\documentstyle[preprint,aps,psfig]{revtex}
\begin{document}
\tightenlines
\draft

\title{
Melting or nucleon transfer in fusion of heavy nuclei ?
}
\author{A. Diaz-Torres$^{1}$, G.G.Adamian$^{1,2,3}$, N.V.Antonenko$^{1,2}$
and W.Scheid$^{1}$}
\address{$^{1}$Institut f\"ur Theoretische Physik der
Justus--Liebig--Universit\"at,
D--35392 Giessen, Germany\\
$^{2}$Joint Institute for Nuclear Research, 141980 Dubna, Russia\\
$^{3}$Institute of Nuclear Physics, 702132 Tashkent, Uzbekistan}
\date{\today}
\maketitle
\title{Melting or nucleon transfer in fusion of heavy nuclei?}

\begin{abstract}
The time-dependent transition between a diabatic interaction potential
in the entrance channel and an adiabatic potential
during the fusion process is investigated within the two-center shell model.
A large hindrance is obtained for the motion to smaller
elongations of near symmetric dinuclear systems.
The comparison of the calculated energy thresholds for
the complete fusion in different relevant collective variables
shows that the dinuclear system prefers to evolve in the mass asymmetry
coordinate by nucleon transfer to the compound nucleus.
\end{abstract}


The experimental synthesis of new superheavy elements \cite{SH,SM}
and nuclei far from the line of stability stimulates
the study of fusion processes in heavy
ion collisions at low energies ($<$ 15 MeV/nucleon).
The experimental results on the fusion of heavy nuclei and
the  cold fusion reactions
can not be satisfactorily explained as a melting of two
nuclei in the relative coordinate with the known macroscopic and microscopic
models \cite{IOC,FR,SW}.
In these models the fusion cross section is overestimated
for near symmetric reactions,
for example,  for the $^{110}$Pd+$^{110}$Pd reaction, and
the optimal bombarding energy and isotopic dependence
of the fusion probabilities are incorrect \cite{SH,SM}.
In the dinuclear system (DNS) concept \cite{V1,V2}
the fusion is explained as a transfer of nucleons or clusters
between nuclei in a dinuclear (quasimolecular) configuration.
Therefore, fusion occurs in the degree of freedom of mass asymmetry
$\eta=(A_2 - A_1)/A$ ($A_1$ and $A_2$
are the mass numbers on both sides of the neck and $A=A_1 + A_2$),
and this model describes the fusion
in symmetric reactions with heavy nuclei
and in reactions producing superheavy nuclei quite well.
These facts stimulate the microscopical justification  of relevant
collective coordinates which are mainly responsible
for the complete fusion.
It will be interesting to
study the competition between two possible fusion channels.
The first one ($\lambda $--channel) describes the transition of two nuclei
into the compound nucleus with the elongation coordinate $\lambda$ of the
two--center shell model (TCSM) which measures the length $l$ of the
two--nucleus system in units of the diameter $2R_0$ of a spherical assumed
compound nucleus: $\lambda=l/2R_0$. This channel assumes a fixed
mass asymmetry during the fusion and is named $\lambda $--channel. The second
channel, named $\eta $--channel, describes the evolution of the DNS
to the compound nucleus as a change of the mass asymmetry $\eta$ by
nucleon transfer from the light nucleus to the heavy one
(the DNS concept \cite{V1,V2}).
The comparison of the fusion probability calculated
in both channels will allow us to find the favorable fusion channel.

In microscopic studies of the dynamics of fusion, based on group theory
and on the TCSM, we obtained a large structural forbiddenness for the
fast growth of the neck  and for the motion to
smaller elongations in the dinuclear system \cite{ST,AAT,aais}.
The effect of structural forbiddenness
is the reason for the stability
of the DNS configurations  against
a dissolution into the $\lambda$--channel.
The decrease of the structural forbiddenness with increasing
mass asymmetry  \cite{AAT} supports the idea of the
DNS concept \cite{V1,V2} that complete fusion takes place after the
mass asymmetry $\eta $ is increased by thermal fluctuations.
In the present letter we will study whether
the system has time for destroying the "memory" on
the structural forbiddenness.
This time is necessary to reorganize the density of the
system for the transition from the initial diabatic potential
$V_{di}(\lambda)$
to the adiabatic potential $V_{ad}(\lambda)$.
The value of $V_{di}(\lambda)$
could be represented
by a frozen density approximation, but
the diabatic potential and the frozen density
potential are conceptually and physically not equivalent
even though the numerical values are similar \cite{Nor10,DAS}.
The repulsive character of the
potential is mainly related to the diabatic particle--hole
excitations  and to the compression effects
in diabatic and frozen density considerations, respectively \cite{Nor10}.
The time dependence of the transition of the potential can be related
to the effective characteristic relaxation time $\tau$ for the shape
degrees of freedom of the system \cite{NR,NGO}
\begin{equation}
V(\lambda,t )=V_{di}(\lambda )\exp (-\int\limits^{t}_0\frac{dt}{\tau
(\lambda,t)})+V_{ad}(\lambda )[1-\exp (-\int\limits^{t}_0\frac{dt}{\tau
(\lambda,t)})].
\label{d1_eq}
\end{equation}
A time-dependent dynamical potential $V(\lambda,t )$
was originally introduced in Refs.\cite{NR,NGO} from
a phenomenological ansatz and applied to study the effects of local
equilibrium in dissipative heavy-ion collisions. The
Eq. (\ref{d1_eq}) may be rewritten as
\begin{eqnarray}
V(\lambda,t ) &=&V_{ad}(\lambda)+\Delta V_{di}(\lambda,t )
\label{d2_eq}
\end{eqnarray}
with $\Delta V_{di}(\lambda,t )=(V_{di}(\lambda)-V_{ad}(\lambda))
\exp (-\int\limits^{t}_0\frac{dt}{\tau(\lambda,t)})$.
The additional part $\Delta V_{di}(\lambda,t)$ can be
microscopically obtained from the diabatic excitation of particle-hole states
\begin{eqnarray}
\Delta V_{di}(\lambda,t )
\approx \sum_{\alpha }\epsilon _{\alpha
}^{di}(\lambda)[n_{\alpha }^{di}(\lambda,t)-n_{\alpha }^{ad}(\lambda)],
\label{d3_eq}
\end{eqnarray}
where the $\epsilon_{\alpha}^{di}(\lambda)$ are diabatic
single--particle energies as a function of the elongation $\lambda$
of the TCSM. The adiabatic occupation numbers
$n_{\alpha }^{ad}(\lambda)$ vary with $\lambda$ according to
a Fermi distribution  with the  temperature
$T(\lambda)=\sqrt{E^*(\lambda)/a}$  ($a=A/12$ MeV$^{-1}$),
where the excitation energy $E^*(\lambda)$ is determined
from total energy conservation.
The exponential factor in (\ref{d1_eq}) is due to the
dependence of the diabatic occupation probabilities
$n_{\alpha }^{di}$ on time expressed by the relaxation
equations \cite{Nor,LCN,N2}
\begin{equation}
\frac{dn_{\alpha }^{di}(\lambda,t)}{dt}=
-\frac{1}{\tau (\lambda,t)}[n_{\alpha
}^{di}(\lambda,t)-n_{\alpha }^{ad}(\lambda)],
\label{occup_eq}
\end{equation}
which is known in the relaxation time approximation.
Due to the residual two-body interactions,
the diabatic occupation probabilities
approach a local (fixed $\lambda$) equilibrium
with an average relaxation time
\begin{equation}
\tau (\lambda,t)=\frac{2\hbar }{<\Gamma(\lambda,t)>}.
\label{tau_eq}
\end{equation}
The factor 2 assumes that two subsequent collisions are
sufficient to establish equilibrium for fixed values of the
collective variable $\lambda$. Here, we
use a minimal value of this
factor (or minimal possible value of $\tau$)
in comparison to Refs. \cite{Nor,LCN,N2}
where this factor was chosen as 3--4.
From the Eq.~(1) it is clear that the  effective
time $\tau$ necessary to reorganized  the densities
of the system corresponds to a mean value of various
relaxation times associated to the shape degrees of freedom of system
which is larger than average single-particle decay time
($\frac{\hbar}{<\Gamma>}$) due to the effect of self--consistency
between collective and  single-particle degrees of freedom \cite{HOFM,LAR}.
The width in Eq.~(\ref{tau_eq})
\begin{equation}
<\Gamma(\lambda,t)>=\sum_{\alpha}\overline{n}^{di}_{\alpha}(\lambda,t)
\Gamma_{\alpha}(\lambda)/{\sum_{\alpha}\overline{n}^{di}_{\alpha}(\lambda,t)}
\label{gamm2_eq}
\end{equation}
is an average width of the particle-states above the Fermi level
($\overline{n}^{di}_{\alpha}=n^{di}_{\alpha}$ for
$\epsilon_{\alpha}^{di}>\epsilon_{F}$) and of the hole-states
under the Fermi level
($\overline{n}^{di}_{\alpha}
=1-n^{di}_{\alpha}$ for $\epsilon_{\alpha}^{di}\leq\epsilon_{F}$).
For the widths $\Gamma _{\alpha}$, the following
expression is used
\begin{equation}
\Gamma _{\alpha }=\Gamma _{0}^{-1}\frac{
(\epsilon _{\alpha }^{di}-\epsilon_{F})^{2}
}{1+[(\epsilon _{\alpha
}^{di}-\epsilon _{F})^{2}]/c^{2}}.
\label{gamm1_eq}
\end{equation}
The parameters $\Gamma _{0}$
and $c$ are known from studies with the optical
model potential and the
effective masses, and their values are in the range
0.030 MeV$^{-1}\leq \Gamma_{0}^{-1}\leq$ 0.061 MeV$^{-1}$
and 15 MeV$\leq c \leq$ 30 MeV.
The results depend weakly on the value of
the parameter $c$.
In the
calculations we take the standard value $c$=20 MeV and consider the cases
with the two extreme values of $\Gamma _{0}^{-1}$
mentioned above.
From an extended expression of Eq.~(7) one can see for very large
free energies $\epsilon _{\alpha }^{di}-\epsilon_{F}$
that the broadening of single-particle
widths due to intrinsic excitation energy of system plays no essential role
in contrast to the case when the excited system
is near the equilibrium state \cite{HOFM,BB,PN}.
Although one may define a local excitation
energy during the decay of the diabatic
potential to the adiabatic one, the concept of temperature is less meaningful
as the system is not locally equilibrated or thermalized \cite{KR}.
More detailed investigations are required to clarify these points.

Since in fusion and quasifission we deal with two-center
systems, we use the two-center shell model (TCSM)
\cite{TC1,TC2,Adeev}
for calculating the adiabatic or diabatic potential energy surface.
In the TCSM the nuclear shapes are defined by a
set of coordinates. The elongation $\lambda$ measures
the length $l$ of the system and is used to describe the relative motion.
The transition of the nucleons through
the neck is described by the
mass asymmetry $\eta$.
The neck parameter $\varepsilon=E_0/E'$ is defined
by the ratio of the actual barrier height $E_0$ to the barrier
height $E'$ of the two-center oscillator.
The deformations $\beta_i=a_i/b_i$ of axial
symmetric fragments are related to the ratio of their semiaxes.

With the TCSM and the Strutinsky method
the adiabatic potential energy $V_{ad}$
can be calculated as
the sum of a macroscopic
energy $U_{LDM}$ obtained by the liquid drop model,
microscopic correction $\delta U_{shell}$
that arises due to the shell
structure of the nuclear system,
energy of the pairing
correction $\delta U_{pair}$ and
the proximity nuclear potential $V_N$ to improve the values of the
adiabatic energy obtained for large elongations \cite{aais,DAS}
\begin{eqnarray}
V_{ad}=U_{LDM}(\lambda,\varepsilon,\eta)+ V_N(\lambda,\varepsilon,\eta)
+\delta U_{shell}(\lambda,\varepsilon,\eta,E^*(\lambda))
+\delta U_{pair}(\lambda,\varepsilon,\eta,E^*(\lambda)).
\label{Pot_eq}
\end{eqnarray}
We neglect the
dependence of potential energy on the angular momentum
in the reactions considered \cite{V2}
because in fusion reactions with heavy nuclei
only low angular momenta ($<20-30\,\hbar$)
contribute \cite{SH,SM,PRL}.
Shell effects are damped exponentially $\delta U_{shell}=\delta
U_{shell}(E^*=0)\exp (-\zeta E^*)$ if
the system is excited with the excitation energy $E^*$. The parameter
$\zeta$ is chosen as
$\zeta ^{-1}=5.48A^{\frac{1}{3}}/(1+1.3A^{-\frac{1}{3}})$ MeV \cite{Ignat}.
The pairing corrections are taken as follows
$\delta U_{pair}$=0 for $E^*\geq E_{c}$ and
$\delta U_{pair}$=$\delta
U_{pair}(E^*=0)[1-E^*/E_{c}]^{2}$ for $E^*<E_{c}$
\cite{SM} with $E_{c}= 10 MeV$.
The isotopic composition of the nuclei forming the DNS was chosen with
the condition of the $N/Z$-equilibrium in the system \cite{V2}.
For the touching configuration of nuclei the
diabatic effects are very small and the diabatic potential practically
coincides with the adiabatic one \cite{DAS}.
So, there is no any difference between  the DNS evolution
in $\eta$ in the adiabatic and diabatic basises.
Because of this the values of fusion barrier in $\eta$--channel and
quasifission barrier are practically independent of
time. For the  smaller elongations $\lambda<\lambda_t$
the diabatic potential is
considerably larger than the adiabatic potential.

The diabatic levels $\epsilon _{\alpha }^{di}$ are
classified by the quantum numbers $\alpha =j_{z},l_{z},s_{z},n_{\rho },n_{z}$
of the eigenstates of the diabatic Hamiltonian $H^{\prime }=H-SV$ which is
obtained by the method of maximum symmetry \cite{LCN,DAS}.
The method of maximum
symmetry eliminates the symmetry violating parts $SV$ from the total
Hamiltonian $H$ of the TCSM.
Diabatic levels obtained by this method agree
with those of the maximum overlap procedure \cite{LCN}. We use the method of
maximum symmetry because it is numerically easy to handle.
In most reactions studied in \cite{DAS}
the diabatic potential has a minimum in the neck parameter
$\varepsilon $ around $\varepsilon$ = 0.65--0.85.
Nuclei are considered as spherical
with $\varepsilon$ = 0.74 which corresponds to
realistic shapes of the DNS for $\lambda$ =1.5--1.6. Due to the
large inertia and friction coefficients in the neck coordinate $\varepsilon$
obtained in
a microscopical treatment \cite{aais}, the
DNS configurations with fixed neck parameters have
a long lifetime in comparison to the reaction time.

In order to study the competition between the fusion channels in $\lambda $
and $\eta$, we use
the fusion rate $\Lambda_{fus}^{\lambda }(t)$ ($\Lambda _{fus}^{\eta }(t)$)
through the inner fusion barrier $B^{\lambda}_{fus}$
($B^{\eta}_{fus}$) in $\lambda$
($\eta $) to calculate the fusion probability in
$\lambda$--channel ($\eta$--channel)
\begin{equation}
P_{fus}^{\lambda (\eta )}=\int\limits_{0}^{t_{0}}\Lambda _{fus}^{\lambda
(\eta )}(t)dt,
\label{probab_eq}
\end{equation}
where the lifetime $t_{0}$ of the DNS is obtained with the condition
\begin{equation}
\int\limits_{0}^{t_{0}}[\Lambda _{fus}^{\lambda }(t)+\Lambda _{fus}^{\eta
}(t)+\Lambda _{qf}^{\lambda }(t)]dt=1.
\label{lifetime_eq}
\end{equation}
The rate of probability $\Lambda _{qf}^{\lambda }(t)$ through
the external barrier in $\lambda$ determines the quasifission process
(the decay of the system).
The height $B_{qf}^{\lambda}$ of this barrier
monotonically decreases with the DNS mass asymmetry (at fixed total mass
and charge numbers of system) because the increase of the Coulomb repulsion
with decreasing $\eta$ leads to very shallow pocket in nucleus--nucleus
potential. So, the quasifission probability for the symmetric and near
symmetric DNS is much larger  than for asymmetric one.
Because the reactions considered in paper are symmetrical or near symmetrical
and, correspondingly, the initial DNS of them are in or near the minimum
of the total potential energy of system as function of $\lambda$ and $\eta$,
the main contribution to quasifission channel comes from the
initial or near initial  configurations which have more or less the same
quasifission barrier $B_{qf}^{\lambda}$. All these facts allow us to
calculate the quasifission rate for the initial DNS  using the Kramers-type
expression. This decay process in $\lambda$  determines mainly
the lifetime of the DNS, because the barrier $B_{qf}^{\lambda}$ is
smaller than the barrier $B_{fus}^\eta$ in mass asymmetry.
The lifetimes $t_{0}$ obtained for the reactions considered
are comparable with the
experimentally extracted characteristic fusion times
of $10^{-21}-10^{-20} s$ \cite{Th} .
Since the
rates assume their final values very fast
\cite{V2,aais} and their initial contributions are of the order of
the accuracy of the calculation of the barrier heights, we use
the one-dimensional
Kramers expression \cite{KRAM}
($^{Kr}\Lambda _{i}^{j}$, $i=$"$fus$" or "$qf$" and
$j=$"$\lambda$" or "$\eta$") which is a quasi--stationary solution
of the Fokker-Planck equation for the corresponding rate of probability
\begin{equation}
^{Kr}\Lambda _{i}^{j}=\frac{\omega_j }{2\pi \omega ^{B^j_{i}}}
\left( \sqrt{(\frac{
g }{2\hbar })^{2}+(\omega ^{B^j_{i}})^{2}}-\frac{g}{2\hbar }\right) \exp
(-\frac{B^j_{i}}{T(\lambda_t)}).
\label{kram1_eq}
\end{equation}
Here, $B^j_i$ denotes the height of the fusion barriers $B^{\lambda}_{fus}(t)$
and $B^{\eta}_{fus}$, and quasifission barrier $B^{\lambda }_{qf}$.
The values of
$B^{\eta}_{fus}$ and  $B^{\lambda }_{qf}$ are practically independent of
time.
The initial DNS of the reactions considered are in or near the minimum
of the total potential energy of system as function of $\lambda$ and $\eta$
as  mentioned above and are in thermodynamic equilibrium because
diabatic and adiabatic potentials practically
coincides. The temperature $T(\lambda_t)$
is calculated by using the expression
$T(\lambda_t)=\sqrt{E^*(\lambda_t)/a}$
where $\lambda_t$ is the elongation
of the touching nuclei, i.e. of the initial DNS.
In the calculations we assume that the excitation energy of the
initial DNS is $E^{*}(\lambda_t)$=30 MeV in all reactions considered.
In (\ref{kram1_eq}),
$\omega^{B^j_i}$ is the frequency of the inverted harmonic oscillator
approximating the potential in the variable $j$ on the
top of the fusion or quasifission barriers  $B^j_i$, and
$\omega_j$ is the frequency of the harmonic oscillator
approximating the potential in the variable $j$ for  the
initial DNS.
The method of the calculation of the mass  parameters
is given in Refs. ~\cite{V2,aais}.

In our calculations
of the fusion probabilities, the following values are used
$\hbar \omega^{B^{\lambda }_{qf}}$ $\approx$ 0.8--1.0 MeV,
$\hbar \omega^{B_{fus}^{\eta }}\approx$ 1.5--2.0 MeV,
$\hbar \omega _{\lambda }\approx$ 1.5--2.0 MeV
and $\hbar \omega_{\eta }\approx$ 0.8--1.0 MeV for the reactions
considered \cite{V2}.
The value of $\hbar \omega^{B_{fus}^{\lambda}}$
at the inner fusion barrier in $\lambda $ is about 0.5--0.6 MeV and agrees
with the one obtained in the calculations of fission \cite{Yama} at the saddle
point. The friction coefficients obtained with $g$=2 MeV have the same
order of magnitude as the ones calculated within the one-body dissipation
models \cite{V2,aais}. The values obtained for
$P^{\lambda(\eta)}_{fus}$ depend rather weakly on $g$ in
(\ref{kram1_eq})\cite{V2}. The possibility to apply the
Kramers expression to relatively small barriers
was demonstrated in \cite{GoKo}.

The time-dependent diabatic potentials for the reactions $^{110}$Pd+$^{110}$Pd
and $^{124}$Sn+$^{124}$Sn are presented in Figs. 1a) and b).
The time-dependent inner fusion barrier $B_{fus}^{\lambda}$ in $\lambda$
appears due to the dependence of the relaxation time of
the diabatic potential on the
elongation $\lambda $ as shown in Fig. 2
for the $^{110}$Pd+$^{110}$Pd reaction. The decrease of $<\Gamma>$
in time causes a slower transition of the diabatic potential
to the adiabatic potential when this potential approaches
the adiabatic limit.
The structures in the vanishing diabatic potential (Fig. 1a)
are caused by the structures in $<\Gamma>$ as a function of
$\lambda$  (Fig.2) which
disappear in time more rapidly than the structures of the diabatic
potential. Fig.~3a shows the dependence of
$B_{fus}^{\lambda }$ on time for
the reactions $^{110}$Pd+$^{110}$Pd ($\eta$=0)
and $^{56}$Cr+$^{164}$Er ($\eta$=0.5)
which produce the same compound nucleus $^{220}$U.
The inner fusion barrier in
$\lambda$ for the asymmetric DNS appears later and
is pronounced smaller than the one for the symmetric DNS and decreases
slower in time. The smaller values of $B_{fus}^{\lambda }$ for
the asymmetric
DNS can be explained by the structural forbiddenness for the
motion to smaller values of $\lambda$ which is smaller than
the hindrance in the
symmetric case \cite{AAT,DAS}.
The lifetime $t_{0}$ of the DNS formed in both reactions
is about $8\cdot 10^{-21}$s and
the values of $B_{fus}^{\lambda}$ at this time are larger than the
corresponding fusion barriers  $B_{fus}^{\eta }$ in $\eta$ (see Fig. 3b and
Table 1). So,
the fusion probability $P_{fus}^{\lambda }$
in $\lambda$ is smaller than  $P_{fus}^{\eta}$ in $\eta$ (Table 2).
This is also demonstrated in Tables 1 and 2
for the reactions $^{123}$Sn+$^{123}$Sn, $^{110}$Pd+$^{136}$Xe,
$^{86}$Kr+$^{160}$Gd and $^{76}$Ge+$^{170}$Er
which lead to the same compound nucleus $^{246}$Fm.
The calculated values of $P_{fus}^{\eta}$  are
in agreement with fusion probabilities extracted
from the experimental data \cite{Ge}.
The fusion barrier along mass asymmetry does not depend on time
because the diabatic potential energy at the touching configurations
with different $\eta$ is very close to the adiabatic potential energy.
From Tables 1 and 2 one can see that
the fusion probability $P_{fus}^{\lambda}$ increases with
increasing mass asymmetry in the entrance
channel.
It follows from our analysis that in the $\lambda$-channel as well as
in the $\eta$-channel the complete fusion in symmetric reactions
yields smaller cross sections
in comparison with asymmetric combinations .

As shown in Figs. 1a) and b), the value of $B_{fus}^{\lambda }(t_{0})$
for the reaction $^{124}$Sn+$^{124}$Sn is larger than the value
for $^{110}$Pd+$^{110}$Pd.
An increasing mass number of the system generally
causes an increase of the
repulsive character of the initial diabatic potential \cite{DAS}
and decreases the  value of the quasifission barrier
which mainly determines the DNS lifetime $t_{0}$.
The same behaviour of
the fusion probability $P_{fus}^{\lambda}$ is
obtained for the symmetric reactions
$^{90}$Zr + $^{90}$Zr, $^{100}$Mo + $^{100}$Mo,
$^{110}$Pd + $^{110}$Pd, $^{123,124}$Sn + $^{123,124}$Sn and
$^{136}$Xe + $^{136}$Xe as well (Table 2).
Despite of the strong decrease of  $B_{fus}^{\lambda}$ with
the change of the parameter $\Gamma_{0}$ from the maximal to
the minimal value,
the fusion probabilities obtained in
the $\lambda$--channel remain to be much
smaller than the fusion probabilities obtained in the $\eta$--channel
which are similar to the experimental values \cite{SM,Ge} (see Table 2).
In the heavier system the difference
between the fusion barriers and probabilities in
both channels is larger and the $\lambda$--channel is practically closed.
This means a dominance of the fusion in the mass asymmetry degree
of freedom which is the fundamental assumption in the DNS concept \cite{V1,V2}.

We discussed which of the two approximations, the diabatic or the adiabatic,
is closer to reality in the fusion process. The actual
situation depends very much on
the initial diabatic potential and on the ratio of the quasifission
time or lifetime $t_0$
of the DNS and the equilibrium time $\tau$.
The time--dependent transition between diabatic
and adiabatic potentials is a slower process
than the quasifission one and the system has not enough time for
destroying the "memory" on
the structural forbiddenness which is in agreement
with estimations in Ref. \cite{AAT}.
As the result, a large hindrance
for the motion to smaller elongations $\lambda$
of the DNS is obtained.
The comparison of the calculated
energy thresholds for the complete
fusion in the $\lambda$-- and $\eta$--channels
shows that the DNS favorably
evolves to the compound nucleus in mass asymmetry
due to the thermal fluctuations.

\vspace{1cm}
We thank Prof. R.V.Jolos, Prof. V.V.Volkov, Prof. Yu. M.Tchuvil'sky,
Prof. P.Hess and Dr. A.B.Larionov
for fruitful discussions. A.D-T. is grateful to the DAAD
for support. G.G.A. thanks the Alexander
von Humboldt-Stiftung for support. This work was
supported in part by DFG and RFBR.

\newpage
\begin{table}[!hbp]
\caption{Quasifission and inner fusion
barriers in $\eta$
and $\lambda$ calculated within the TCSM
for various symmetric and asymmetric reactions.
The inner fusion barriers in
$\lambda$ are given for the lifetimes $t_0$ of the DNS
formed in these reactions.
The notations 1) and 2) mean that  the values of $B_{fus}^ {\lambda} (t_0)$
are calculated with $\Gamma_{0}^ {-1}$= 0.030 MeV$^ {-1}$
and 0.061 MeV$^ {-1}$,
respectively.}
\begin{tabular}{|c|c|c|c|c|c|}
\hline
Reactions & $B_{qf}^{\lambda }$ & $B_{fus}^{\eta }$ & $t_{0}$ & $^{1)}$
$B_{fus}^{\lambda }(t_0)$ & $^{2)}$ $B_{fus}^{\lambda }(t_0)$ \\
& [MeV] & [MeV] & $[10^{-21}$ s] & [MeV] & [MeV] \\ \hline
 $^{90}$Zr+$^{90}$Zr  $\to ^{180}$Hg & 2.9 & 6 & 20 & 10 & 4 \\
$^{100}$Mo+$^{100}$Mo $\to ^{200}$Po & 2.2 & 8 & 15 & 12 & 5 \\
$^{110}$Pd+$^{110}$Pd $\to ^{220}$U  & 1.3 & 12 & 8 & 36 & 14 \\
 $^{56}$Cr+$^{164}$Er $\to ^{220}$U  & 2.6 & 2  & 8 & 14 & 4 \\
 $^{76}$Ge+$^{170}$Er $\to ^{246}$Fm & 0.4 & 10 & 5 & 53 & 27 \\
 $^{86}$Kr+$^{160}$Gd $\to ^{246}$Fm & 0.2 & 12 & 4 & 65 & 39 \\
$^{110}$Pd+$^{136}$Xe $\to ^{246}$Fm & 0.1 & 15 & 3 & 91 & 54 \\
$^{123}$Sn+$^{123}$Sn $\to ^{246}$Fm & 0.1 & 16 & 3 & 112 & 67 \\
$^{136}$Xe+$^{136}$Xe $\to ^{272}$Hs & 0 & 22 & 2 & 237 & 154 \\
\hline
\end{tabular}
\end{table}
\newpage

\begin{table}[!hbp]
\caption{Fusion
probabilities $P_{fus}^{\lambda,\eta }$ in the $\lambda$-- and $\eta$--
channels calculated for the reactions presented in the Table 1 are compared
with known experimental values $P_{fus}^{\exp}$
\protect\cite{SM,V2,aais,Ge}.
The notations 1) and 2) are the same as in
Table 1.   }
\begin{tabular}{|c|c|c|c|c|}
\hline
Reactions & $^{1)}$ $P_{fus}^{\lambda }$ & $^{2)}$ $P_{fus}^{\lambda }$ &
$P_{fus}^{\eta }$ & $P_{fus}^{\exp}$ \\
\hline
$^{90}$Zr+$^{90}$Zr $\to ^{180}$Hg   & 2$\cdot $10$^{-4}$ & 2$\cdot $10$^{-2}$
& 2$\cdot $10$^{-1}$  & $\sim $10$^{-1}$ \\
$^{100}$Mo+$^{100}$Mo $\to ^{200}$Po & 9$\cdot $10$^{-6}$ & 3$\cdot $10$^{-3}$
& 2$\cdot $10$^{-2}$  & 5$\cdot $10$^{-2}$ \\
$^{110}$Pd+$^{110}$Pd $\to ^{220}$U & 7$\cdot $10$^{-15}$ & 4$\cdot $10$^{-7}$
& 3$\cdot $10$^{-4}$  & $\sim $10$^{-4}$ \\
$^{56}$Cr+$^{164}$Er $\to ^{220}$U  &1$\cdot $10$^{-6}$ & 2$\cdot $10$^{-3}$
& 6$\cdot $10$^{-1}$  & \\
$^{76}$Ge+$^{170}$Er $\to ^{246}$Fm  & 9$\cdot $10$^{-22}$ & 3$\cdot $10$^{-12}$
& 6$\cdot $10$^{-4}$  & 8$\cdot $10$^{-4}$ \\
$^{86}$Kr+$^{160}$Gd $\to ^{246}$Fm  & 4$\cdot $10$^{-26}$ & 2$\cdot $10$^{-16}$
& 7$\cdot $10$^{-5}$  & 5$\cdot $10$^{-5}$ \\
\hline
\end{tabular}
\end{table}

\newpage

\begin{figure}
\psfig{figure=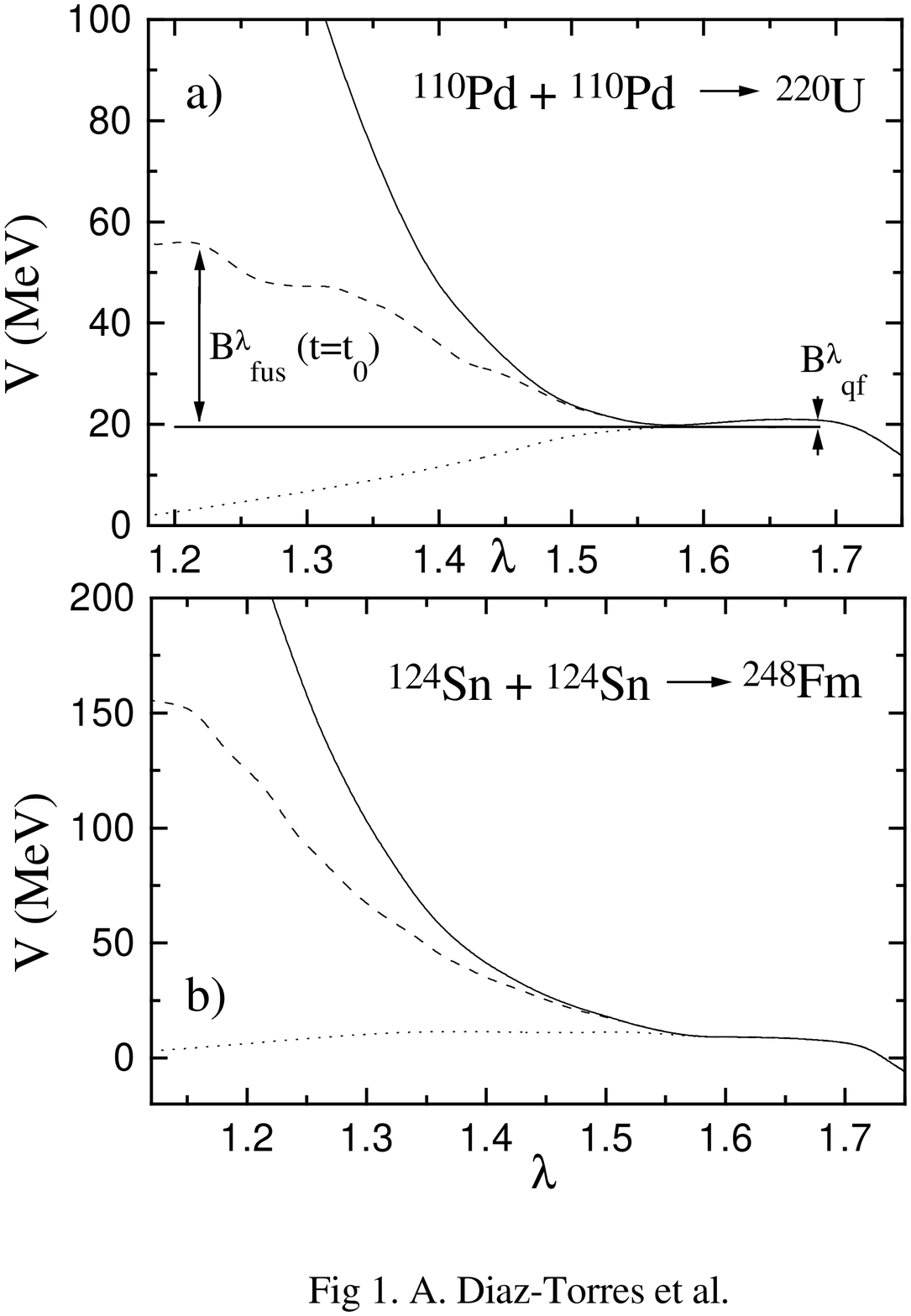,width=14cm,height=16.0cm}
\caption{a) Time-dependent dynamical potential $V(\lambda,t)$
as a function of
elongation $\lambda$ for the system $^{110}$Pd+$^{110}$Pd.
The initial diabatic potential $V(\lambda,t=0)=V_{di}(\lambda)$
and the adiabatic potential $V_{ad}(\lambda)$
are shown by solid and dotted curves, respectively.
The diabatic potential $V(\lambda,t=t_0)$
at the lifetime $t_0$ of the DNS
is presented by a dashed curve.
The nuclei are considered spherical with the neck parameter
$\varepsilon=0.74$.
The parameter $\Gamma_{0}^{-1}= 0.030$ MeV$^ {-1}$
is used in the calculation of the single--particle widths.
The fusion $B^{\lambda}_{fus}(t=t_0)$
and quasifission $B^{\lambda}_{qf}$
barriers  in $\lambda$ are indicated.
These barriers are measured with respect to the minimum of the 
potential. b) The same as in a) but for the system $^{124}$Sn+$^{124}$Sn.
}
\label{1_fig}
\end{figure}
\newpage

\begin{figure}
\psfig{figure=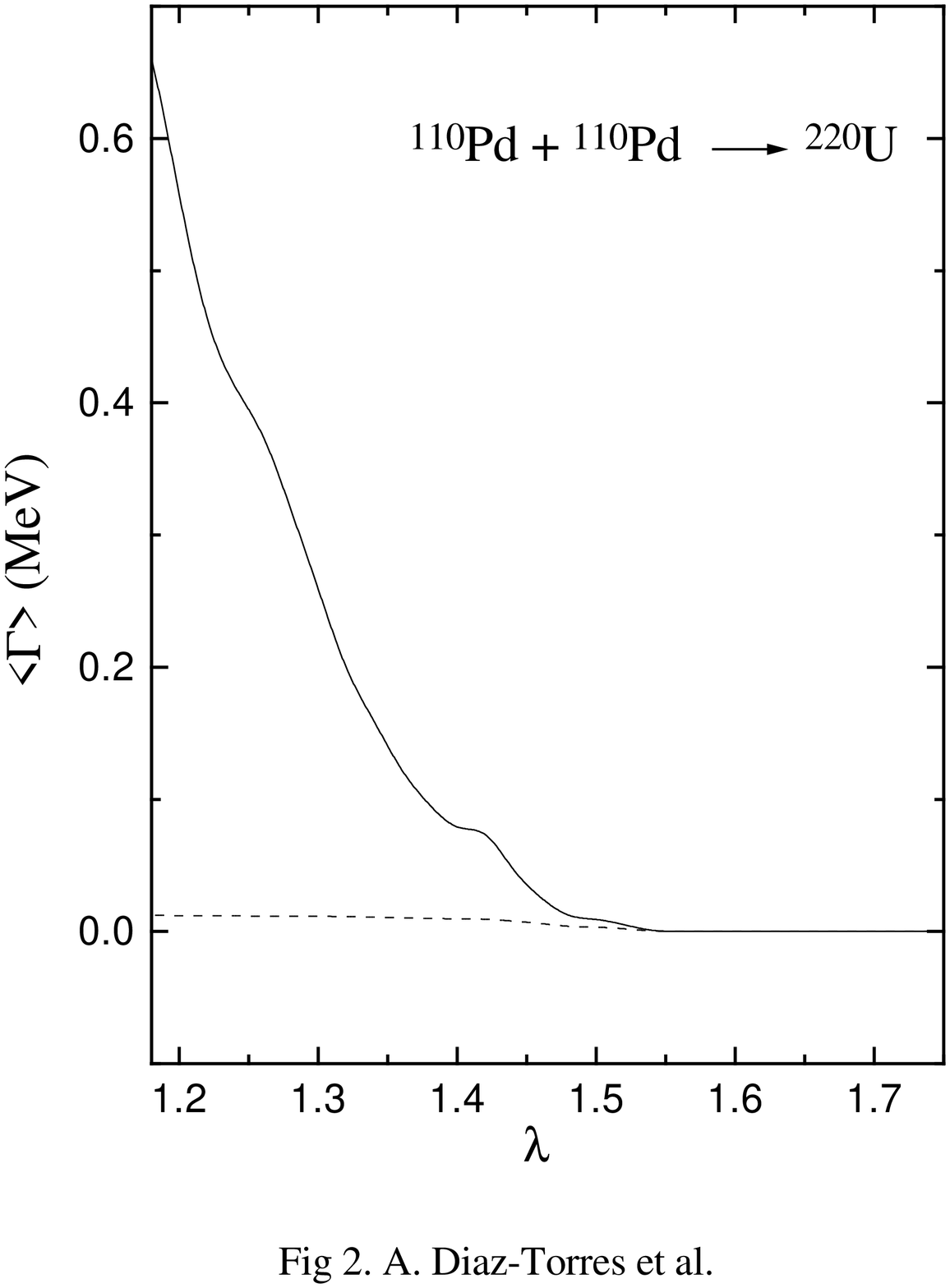,width=14cm}
\caption{ Time-dependent  average width $<\Gamma(\lambda,t)>$ as a function
of elongation $\lambda$ for the system $^{110}$Pd+$^{110}$Pd.
The dependences $<\Gamma(\lambda,t=0)>$  and $<\Gamma(\lambda,t=t_0)>$
are shown by solid and dashed curves, respectively.}
\label{2_fig}
\end{figure}
\newpage

\begin{figure}
\psfig{figure=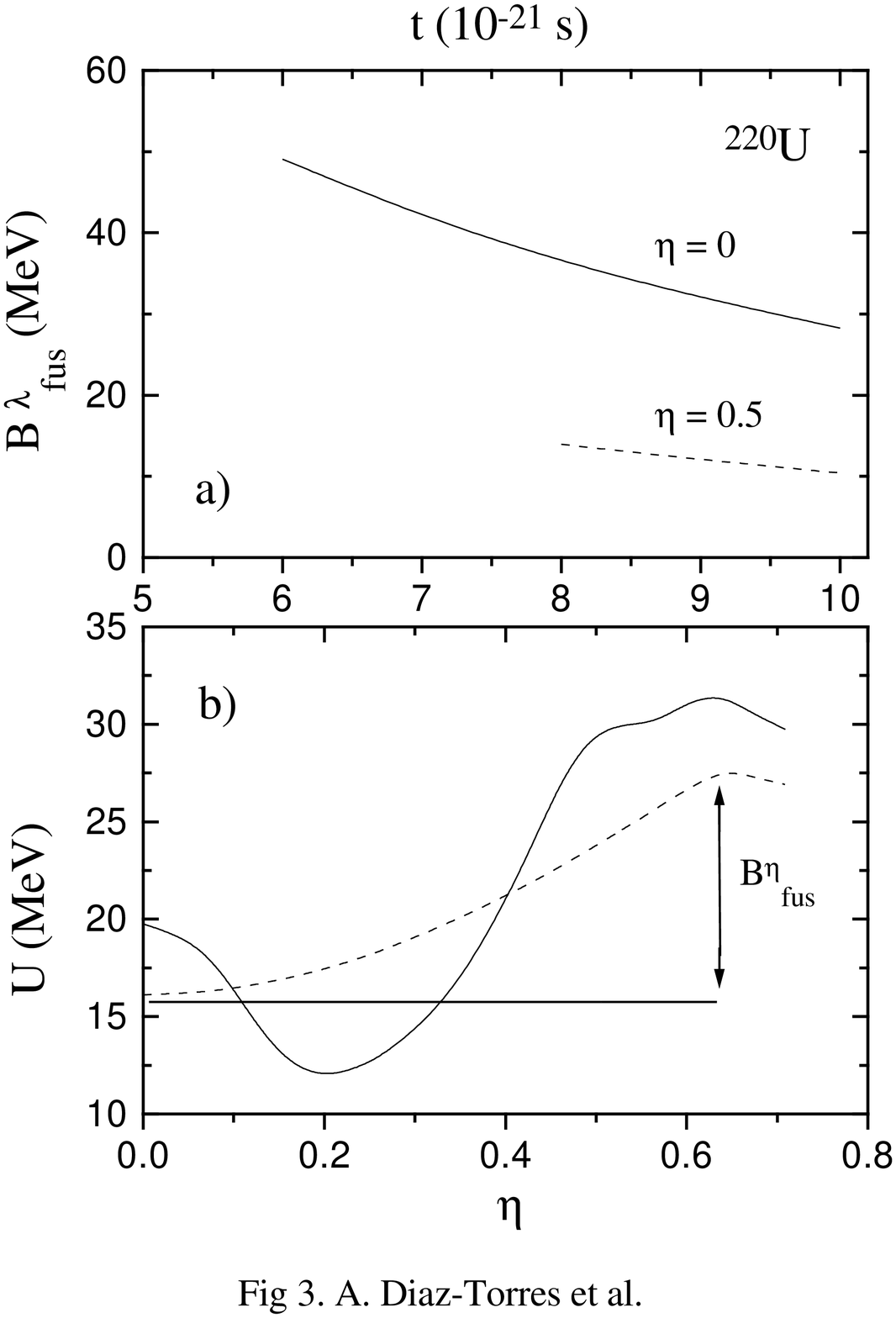,width=14cm,height=15.5cm}
\caption{a) Inner fusion barriers
$B^{\lambda}_{fus}(t)$ in $\lambda$ as a function of time for the systems
$^{110}$Pd+$^{110}$Pd (solid curve) and
$^{56}$Cr+$^{164}$Er (dashed curve) which produce the same
compound nucleus $^{220}$U.
The nuclei are considered spherical with $\varepsilon=0.74$.
The parameter $\Gamma_{0}^ {-1}$ = 0.030 MeV$^ {-1}$ is used
in the calculation of single--particle widths.
For times smaller than 6$\times 10^{-21}$ s and 8$\times 10^{-21}$ s for
$\eta$=0 and 0.5, respectively, the potential
$V(\lambda ,t)$ is only repulsive and has no barrier (see Fig.~1).
b) Calculated adiabatic potential energy of the DNS
in the touching configuration of the nuclei
as a function of mass asymmetry $\eta$
for reactions leading to the same compound nucleus $^{220}$U.
The potential is calculated  within
the adiabatic TCSM with (solid curve) and without (dashed curve)
shell corrections.
The fusion barrier $B^{\eta}_{fus}$ in $\eta$ for the system
$^{110}$Pd+$^{110}$Pd is shown.}
\label{3_fig}
\end{figure}

\end{document}